# Conceptual Design of Human-Drone Communication in Collaborative Environments

Hans Dermot Doran
*Inst. of Embedded Systems / ZHAW*
Winterthur, Switzerland
donn@zhaw.ch

Monika Reif
*Inst. of Applied Mathematics and Physics / ZHAW*
Winterthur, Switzerland
reif@zhaw.ch

Marco Oehler
*Zurich University of Applied Sciences*
Winterthur, Switzerland
oehlemar@students.zhaw.ch

Curdin Stöhr
*Zurich University of Applied Sciences*
Winterthur, Switzerland
stoehcur@students.zhaw.ch

Pierluigi Capone
*Centre for Aviation / ZHAW*
Winterthur, Switzerland
capo@zhaw.ch

*Abstract* - **Autonomous robots and drones will work collaboratively and cooperatively in tomorrow's industry and agriculture. Before this becomes a reality, some form of standardised communication between man and machine must be established that specifically facilitates communication between autonomous machines and both trained and un-trained human actors in the working environment. We present preliminary results on a human-drone and a drone-human language situated in the agricultural industry where interactions with trained and un-trained workers and visitors can be expected. We present basic visual indicators enhanced with flight patterns for drone-human interaction and human signaling based on aircraft marshalling for humane-drone interaction. We discuss preliminary results on image recognition and future work.**

*Keywords—unmanned aerial vehicles, autonomous drones, collaborative drones, human drone communication, drone marshalling, symbolic aggregate approximation.*

## I. Introduction

Concurrent with the increase of drones in real-world applications, questions on safe interaction between human collaborators and low-cost drones are not being answered. There is research available on embodied emotional communication [1, 2], natural interaction techniques [3] and a plethora of techniques using non-visual or electronically-assisted techniques [4]. There is also a considerable body of work in pose and gesture recognition generally which tend towards using interesting algorithmic techniques like neural networks and/or relatively expensive and power-hungry sensory systems like the Kinect [5-8]. These methods unfortunately do not appear to promise rapid passage through relevant safety certification before deployment, given it can be fully expected that safety in drone collaboration applications will be become an urgent requirement.

In order for drones to be considered for applications requiring adherence to safety standards, certain characteristics are required including acknowledgement of known regulations and focus on robustness of technologies. In this body of work, we investigate the application of human drone communication using low-cost drones.

We pick as our use case a known issue namely drones sharing workspace with humans in cherry plantations where the drones collect data from fly traps [9] which indicate whether further action, for instance spraying, needs to take place. Given that this data collection will occur in the presence of humans who may be blocking access to the fly traps, a negotiated access to the traps must take place. The rest of the paper is structured accordingly. In the next section we derive the requirements from user stories, we discuss some practical factors in the communication between drones and humans and vice versa. In Section III we present some suggestions and preliminary results which we discuss in Section VI before suggesting further work.

## II. Requirements Derivation

We largely assembled the relevant requirements via the creation of user-stories based around three characters, orchard supervisor, orchard worker and orchard visitor, corresponding roughly to well trained, partially trained and non-trained persons in collaborative activities with drones. These user stories – narrative building as understood by early agile development systems [10] rather than the current formulistic approach [11] – resulted in a set of minimum communication requirements between both drones and collaborators and vice versa [12].

Amongst the primary requirements are an indication of which horizontal direction the drone is flying so, based on FAA regulations [13], a ring with 10 tri-colour light emitting diodes was constructed and attached to the experimental drone (Figure 1)

[14]. Depending on the direction of controlled flight, the position of red, green and white lighting will change. Power requirements with respect to illumination distance is an issue that needs further consideration. There is obvious scope for optimisation by the use of separate high luminosity LEDs. The integration of an appropriate sensor like an IMU to indicate actual flight is yet to be discussed in greater detail.

The ring can be turned to all red should a safety function be triggered, which can be achieved as a default setting [15]. There was no consensus on whether an all-green ring would find application.

An additional, vertical, LED array was added to indicate whether the drone was taking off (see legs of the drone - animation from bottom to top) or landing (top to bottom) but user-feedback indicated that they are difficult to distinguish, do not serve clarity, indeed serve to confuse, and so will be discarded in future versions. Since in vertical take-off/landing situations directional lights are not necessary, a combination of RGB light signals may be used to indicate these flight patterns, this is left for further work.

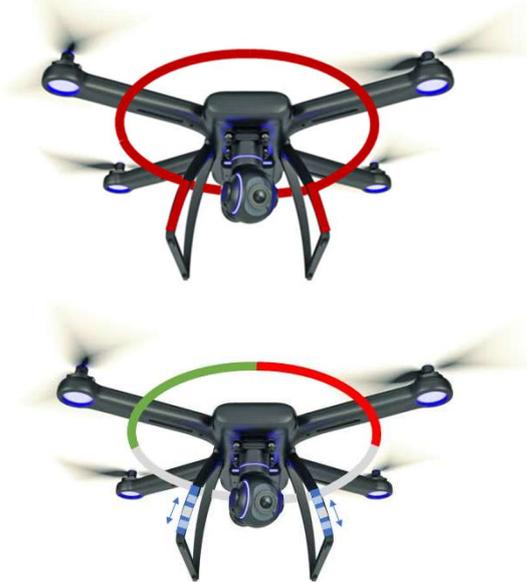

*Figure 1: Top: Drone All-Round-Light Switched to Danger (Red). Bottom: All-Round-Lights Switched to Navigation*

### III. FLIGHT PATTERNS AND MARSHALLING SIGNAGE

Both collaborators and observers can be expected to appreciate defined, observable and reproducible behavioural patterns which become a facet of embodied communication [16]. Three standard flight patterns and four communicative flight patterns were identified and/or defined. Standard flight are take-off, landing and actual flight which in our conception are vertical lift-off to flying height, horizontal flight and vertical landing. In addition a "poke" to attract attention, a nodding and a turning to indicate yes and no respectively and a pattern to indicate that the drone wishes to enter the area covered by the person were also defined. These patterns are directly derived from the user stories and are not to be considered a final suggestion.

The standard flight patterns (Figure 2) make sense in terms of flight at a safe height and defined landing and take-off patterns that only vary if the drone is somehow defective or, for instance, caught in wind gusts. The communicative flight patterns are unmistakable flight patterns and thus can be considered an embodied statement of intent by the drone.

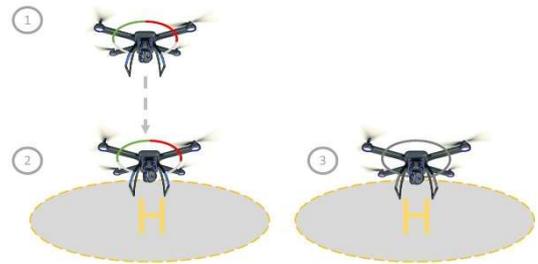

*Figure 2: Landing Flight Pattern. 1: The drone reduces altitude until landed (2) and 3: Once the rotors are switched off the navigation lights are extinguished:*

One of the simple, time-honoured and robust all-weather communication-by-sight forms is by semaphore [17] or marshalling signs [18], which we utilise here. Our use-case demands simple signalling without the use of flags, paddles and/or lights. The use case also demands simple signals that are quickly learnable by disinterested humans. The signage must also be such that it is simply and robustly detectable by low-cost drones, preferably without, or with minimal, addition of an extra camera or sensors. It is completely plausible that applications with more sophisticated modes of collaboration may require more sophisticated signage. Efficiency considerations determine that rather that every drone understand all signs, cost-efficient drones need only understand the bare minimum of signs and so reduce the complexity and cost of recognition electronics. With such an approach we avoid the expense and inconvenience of special wearable equipment [19].

The signage discussed here will reflect what we believe is a minimum necessary set and, considering the robustness requirement, of easily identifiable signs and we specify three static signs.

In accordance with our user stories, the drone will approach the human collaborator and once at the boundaries of a safe distance will "poke" the collaborator to gain the collaborators attention. It is expected that both the visual and the acoustics will alert the collaborator who has the choice of ignoring the approach or responding. In our vision, the collaborator responds with an "attention gained" sign (Figure 3,) after which communication between the two can proceed. Also according to the user stories, the drone will then fly a pattern indicating it wishes to occupy the space where the collaborator is which we have defined as a flying a rectangle to signify area. The two possible answers here are

"Yes" and "No" which we have modelled after are well-known (Switzerland) emergency services signs (Figure 3)

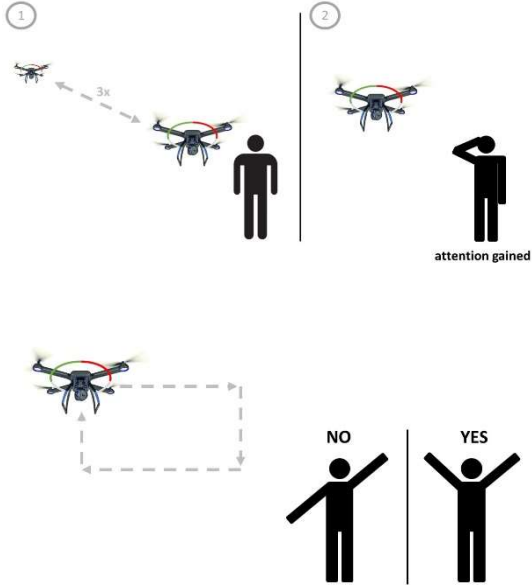

*Figure 3: The drone flies a pattern indicating it wishes to occupy the space currently occupied by the human collaborator. The human collaborator responds with either a Yes or a No*

The "attention gained" sign may be understood as a human-reflex sign to an approaching danger emulating a person putting their hand up to protect their face. In particular it is also unique vis-a-vis known (Swiss) marshalling helicopter signs [20]. In sum, as simple signs that many people with a bare minimum of training will be able to remember, real-world use will require a safe, robust and fast electronic recognition of these signs.

### IV. IMPLEMENTATION NOTES

The recognition algorithm must be rotation invariant as the drone will not be stationary vis-à-vis its communication partner but may need to move into a better position. Recognition must be achieved within real-time constraints. To achieve both rotation invariance and real-time performance we have made a preliminary investigation into using the data-mining algorithm SAX [21] which is, to the best of our knowledge, the first published use of this technique in real-time vision recognition. Briefly, this technique includes converting shapes into a time-series, standardising this time series, apply piecewise aggregation to reduce dimensionality and converting the aggregate to a string of characters. This last step facilitates a comparison of the string against a database of strings and hence can be used quite effectively to identify features in images. Whilst the pre-processing of the image, the conversion of the image into a standardised time-series initially appears expensive, the computational effort for dimension reduction, conversion into a string and the string search are considered computationally cheap.

One example of the sign "No" is shown in Figure 4 with the drone at an altitude of five meters, three meters distance from the signaller, at two (relative azimuth) orientations with respect to the signaller, full-on (0°) and at 65°. Using the 0° relative azimuth image as the canonical reference, the current SAX implementation identifies the "No" sign at altitudes from 2 m to 5 meters (at 3 meters horizontal distance.) At relative azimuth angle's greater than 65°, even with tuning of the piecewise aggregation and alphabet size [22] recognition appears erratic. This result implies that there is a dead angle of 100° where this sign cannot be recognised. The produced SAX string in those dead angles does not, unfortunately, lead us to believe that the drone can use this string as an indicator of which direction to fly in to improve its positioning such that a sign can be recognised. The figure also shows the time-series for each variant of the sign.

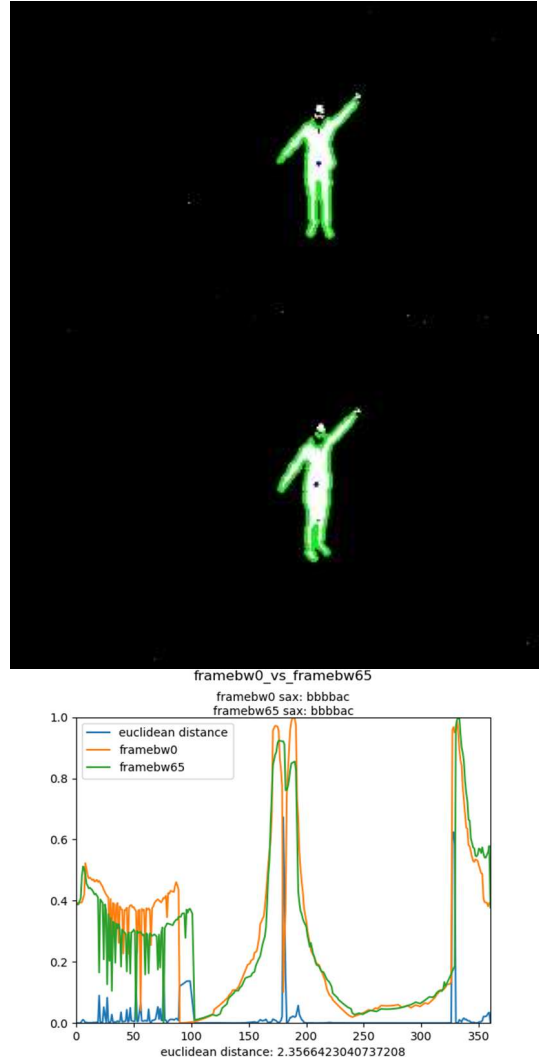

*Figure 4: Top "No" Sign at relative azimuth 0°. Middle relative azimuth 65°. Bottom comparison time series for the two relative azimuths (framebw0 and framebw65 resp.)*

In this configuration the detection of the "No" sign, on a PC (Intel i7 – 7660 U, 2.5 GHz, Windows 10) with un-optimised code written in python (3.7.4 64bit Anaconda distribution) and using openCV (version 4.2) functionality, recognition times for [0°, 65°] are 38 ms and 27 ms respectively. We believe there is enough scope for optimised bare-metal C code to easily achieve 30 frames-per-second (fps) and, with hardware offloading, under 60 fps. Preliminary results also suggest that the strings retrievable from the three signs are unique.

## V. Conclusions

Whilst some authors, [23, 24], have concentrated on close integration of collaborators and drones we have focused on the fundamental safety aspects first. We have co-opted a simple marshalling signage system, suggested a communication processes between man and drone in a collaborative use-case and presented preliminary results that suggest the system could be integrated into performance constrained hardware for use in a low-cost drone using computationally-cheap image recognition algorithms.

There is still a considerable amount of work to be done. Current work is focused on showing the uniqueness of sign identification and a better estimation of the expected real-time characteristics. A complete re-write of the code base according to state-of-the-art high integrity coding and architecture expectations must follow at a later stage. The flexibility of the system with respect to other static and, possibly later, dynamic marshalling signals should also be examined as it is foreseeable that such a system can be used with drones in more complex applications.